\RequirePackage{fix-cm}
\documentclass[twocolumn]{svjour3}          
\smartqed  
\usepackage{graphicx}
\usepackage{acronym}
\usepackage{subcaption}
\usepackage{pgfplots}
\pgfplotsset{compat=1.17}
\usepackage{listings}
\usepackage{amsmath}
\usepackage{amsfonts}
\usepackage[backend=biber]{biblatex}

\bibliography{sample}

\usepackage{showexpl}
\newcommand{\range}[3]{#1\in[#2,...,#3]}

\newacro{fft}[FFT]{fast Fourier transform}
\newacro{dft}[DFT]{discrete Fourier transform}
\newacro{dct}[DCT]{discrete cosine transform}
\newacro{dit}[DIT]{decimation-in-time}
\newacro{dif}[DIF]{decimation-in-frequency}
\newacro{dma}[DMA]{direct memory access}
\newacro{hls}[HLS]{high-level synthesis}
\newacro{rtl}[RTL]{register transfer level}
\newacro{fpga}[FPGA]{field-programmable gate array}
\newacro{asic}[ASIC]{application specific integrated circuit}
\newacro{ip}[IP]{integrated processor}
\newacro{fp}[FP]{floating point}
\newacro{fxp}[FxP]{fixed point}
\newacro{bfp}[BFP]{block floating point}
\newacro{pe}[PE]{processing element}
\newacro{mdc}[MDC]{multi-path delay commutator}
\newacro{sdc}[SDC]{single-path delay commutator}
\newacro{sdf}[SDF]{single-path delay feedback}
\newacro{snr}[SNR]{signal-to-noise ratio}
\newacro{ofdm}[OFDM]{orthogonal frequency-division multiplexing}
\newacro{dsp}[DSP]{digital signal processing}
\newacro{cem}[CEM]{constrained energy minimization}
\newacro{fts}[FTS]{Fourier Transform Spectrometer}
\newacro{svd}[SVD]{singular value decomposition}
\newacro{tsvd}[TSVD]{truncated singular value decomposition}
\newacro{pinv}[PINV]{pseudo-inverse}
\newacro{tik}[TIK]{Tikhonov regularization}
\newacro{fts}[FTS]{Fourier transform spectrometer}
\newacro{opd}[OPD]{optical path difference}
\newacro{sv}[s.v.]{singular value}

\begin{document}

\title{
	Implementation of Hyperspectral Inversion Algorithms on FPGA: Hardware Comparison using High Level Synthesis
	\thanks{This work is partly supported by grant ANR FuMultiSPOC (ANR-20-ASTR-0006) and by AuRA region and FEDER (ImSPOC-UV: convention FEDER n. RA0022348).}
}



\author{El Mehdi Abdali \and
        Daniele Picone \and
        Mauro Dalla Mura \and
        St\'{e}phane Mancini 
}


\institute{
	E. M. Abdali \at
    Univ. Grenoble Alpes, CNRS, Grenoble INP, TIMA, 38031 Grenoble, France \\
    \email{elmehdi.abdali@gmail.com}
    \and
    D. Picone \at
    Univ. Grenoble Alpes, CNRS, Grenoble INP, GIPSA-lab, 38000 Grenoble, France \\
    \email{daniele.picone@grenoble-inp.fr}
    \and
    M. Dalla Mura \at
    Univ. Grenoble Alpes, CNRS, Grenoble INP, GIPSA-lab, 38000 Grenoble, France\\
    Institut Universitaire de France (IUF), 75005 Paris, France
    \\
    \email{mauro.dalla-mura@grenoble-inp.fr}
    \and
    S. Mancini \at
    Univ. Grenoble Alpes, CNRS, Grenoble INP, TIMA, 38031 Grenoble, France \\
    \email{stephane.mancini@univ-grenoble-alpes.fr}
}

\date{Received: date / Accepted: date}

\maketitle

\begin{abstract}
Hyperspectral imaging is gathering significant attention due to its potential in various domains such as geology, agriculture, ecology, and surveillance. However, the associated processing algorithms, which are essential for enhancing output quality and extracting relevant information, are often computationally intensive and have to deal with substantial data volumes.

Our focus lies on reconfigurable hardware, particularly recent FPGAs. While FPGA design can be complex, High Level Synthesis (HLS) workflows have emerged as a solution, abstracting low-level design intricacies and enhancing productivity.

Despite successful prior efforts using HLS for hyperspectral imaging acceleration, we lack a comprehensive research to benchmark various algorithms and architectures within a unified framework.

This study aims to quantitatively evaluate performance across different inversion algorithms and design architectures, providing insights for optimal trade-offs for specific applications. We apply this analysis to the case study of spectrum reconstruction processed from interferometric acquisitions taken by Fourier transform spectrometers.

\keywords{Hyperspectral imaging \and Reconfigurable hardware \and Inversion algorithms \and High level synthesis}
\end{abstract}

\section{Introduction}
\label{sec:intro}



Hyperspectral imaging has gained considerable interest both in the industry and the scientific community for its promising capabilities in a wide range of fields, such as geology, agriculture, ecology and  surveillance~\cite{carrasco03hyperspectral, serranti15hyperspectral, manolakis2016hyperspectral}.
However, the associated processing algorithms needed to improve the quality of the user-end product or to extract information of interest from the captured data are generally computation-intensive~\cite{lopez2013promise} and run on large data volumes~\cite{lei2018deep}. These issues are becoming even more relevant in more recent applications, as the dimensionality of hyperspectral data has increased with finer temporal resolutions made available by novel technologies~\cite{lopez2013promise}. Therefore, fast and efficient computation is required to perform the necessary processing~\cite{Rud06, Wu18, Haut18, Pan18, Fath19}, especially for onboard applications.
In this study, we focus our attention over reconfigurable hardware; these devices have became a preferred target for \ac{dsp} applications, as they can be more easily mass-produced in comparison to equivalent efficient \ac{asic} implementations. In particular, recent \acp{fpga} show increased hardware capabilities at an incrementally decreased cost, making them competitive for many applications that were originally implemented over \acp{asic}~\cite{uzun2005fpga}.

However, using \acp{fpga} architectures with low level design flows, such as manually coded \acp{rtl}, significantly decreases the designer’s productivity, as they prevent an exhaustive exploration of the design space. Moreover, \acp{rtl} are usually technology-dependent, which makes \acp{ip} difficulty portable \cite{fingeroff2010high}. All these reasons have made it necessary to adopt \ac{hls} design workflows that keep up with the design complexity \cite{cong2011high}. Abstracting the architecture of \ac{rtl} can in fact vastly speed-up the design process, allowing to explore multiple design choices (i.e. parallelism, frequencies) to make reasonable trade-offs. By choosing the appropriate optimizations in the most advanced \ac{hls} tools, hardware designers can generate circuits with a precise control of the performance-resource trade-off while ensuring a high productivity \cite{nane2015survey}.
\Ac{hls} tools have the advantage to allow the designer to focus on the algorithms and automatically infer the low-level hardware architecture, while also giving him the possibility to drive low-level features when mandatory.

Despite the promising results demonstrated by previous studies in using \ac{hls} for accelerating hyperspectral imaging algorithms~\cite{altamimi2021systematic}, there is still a lack of a comprehensive research that benchmarks different algorithms and architectures using a unified framework. 
%
Domingo et al.~\cite{domingo2017high} evaluated the use of Intel OpenCL SDK for accelerating spectral classification in hyperspectral imaging using K-Nearest Neighbour as a test case with several optimization techniques, including loop unrolling,  kernel vectorization, and bitwidth optimization.
%
Lei et al \cite{lei2018deep}  proposed a deep pipelined structure for a subpixel target detection algorithm, called \ac{cem}, and improved the efficiency of the algorithm by breaking the computation into multiple stages to achieve higher pipeline acceleration.
%
\cite{gonzalez2011fpga} proposed an \ac{rtl}-based implementation of the hyperspectral N-FINDR, a data analysis algorithm able to extract subpixel information on the materials present in the scene. 
The authors optimized the performance by including a \ac{dma} module and using a pre-fetching technique to hide the input/output interfacing latency.

In this work, we aim to provide a quantitative evaluation of the performance of different inversion algorithms and design architectures, so the most suitable trade-off can be identified for a given application and design a hardware prototype that meets the performance and cost requirements. 

The selection of a proper inversion algorithm is particularly crucial for hyperspectral image reconstruction, where the development of custom hardware solutions can be time-consuming and costly.
We tackle these problems in this work by analyzing a specific case study of spectrum reconstruction for interferometric acquisitions taken by \acp{fts}~\cite{Grif07}. For such devices, for which the Michelson interferometer is the most well-known example, the acquisition, known as \textit{interferogram}, is captured in the Fourier domain instead of the original one. A straightforward strategy to recover the information in the original domain consists in taking the \ac{dct} of the acquisition. While efficiently implemented on hardware~\cite{Sada22} and through \ac{fft}~\cite{makhoul1980fast}, various limitations apply in practical scenarios. For example, the \ac{fft} requires the interferogram to be regularly sampled, and most importantly, the Fourier transform often does not accurately describe the transfer function of many \acp{fts}, such as in the case of the Fabry-Perot interferometer.
For such reasons, various alternative solutions have been proposed in the literature to deal more efficiently with such reconstructions. Some widespread examples, such as those based on penalized \ac{svd}~\cite{Hans90, Tikh95}, do not have equally commonplace hardware implementations in \acp{fpga} or \ac{dsp}.

The main contribution of this paper is to provide a hardware-centric implementation benchmark of various inversion methods for hyperspectral image reconstruction.
This benchmark is motivated by the difficulty to explore large design space and to be able to choose the appropriate method that fits the requirements as well as the constraints. The novel content of our proposed benchmark can be summarized as follows:
\begin{itemize}

	\item We propose an improvement of the Coley-Tukey implementation of the \ac{fft} with a pre- and post-processing normalization step for each computational block. This stage allows to share the exponent of the \ac{bfp} between each \ac{fft} stage, resulting in accelerated calculation of about 3 times.
        
	\item We conduct a hardware comparison of different hyperspectral reconstruction algorithms by evaluating their resource overhead and acceleration gain. This comparison can help establish a new criterion for choosing between these methods, with easily interpretable benefits and limitations to their extension to real-world use cases.
	
	\item We investigate the impact of arithmetic approximations on the quality of the hyperspectral reconstruction. We analyze the fixed-point data width and its effects on both reconstruction quality and hardware resource utilization. By highlighting the impact of arithmetic approximations on reconstruction quality, this research can help identifying the optimal trade-off between quality and efficiency for a given hyperspectral imaging application.
\end{itemize}

The remainder of the paper is organized as follows: in Section~\ref{sec:problem} we describe the case study and some of the algorithmic implementations used to solve it, in Section~\ref{sec:hardware} we discuss their hardware implementations and the relevant criteria to compare their performances, and in Section~\ref{sec:experiments} we describe the related experiments.

\section{Spectrum reconstruction techniques}
\label{sec:problem}

In this section we address the problem of spectrum reconstruction from the acquisition of \acp{fts} by describing some classes of available approaches from the literature. In particular, Section~\ref{ssec:problem_fts} describes the optical theory behind the formation of an interferogram, Section~\ref{ssec:problem_fft} describes the procedure of spectrum recovery through Fourier inspired techniques, while Section~\ref{ssec:problem_svd} describes those based on the \ac{svd}.

\subsection{Problem statement}
\label{ssec:problem_fts}

\begin{figure}
    \centering
    \includegraphics[width=0.9\linewidth]{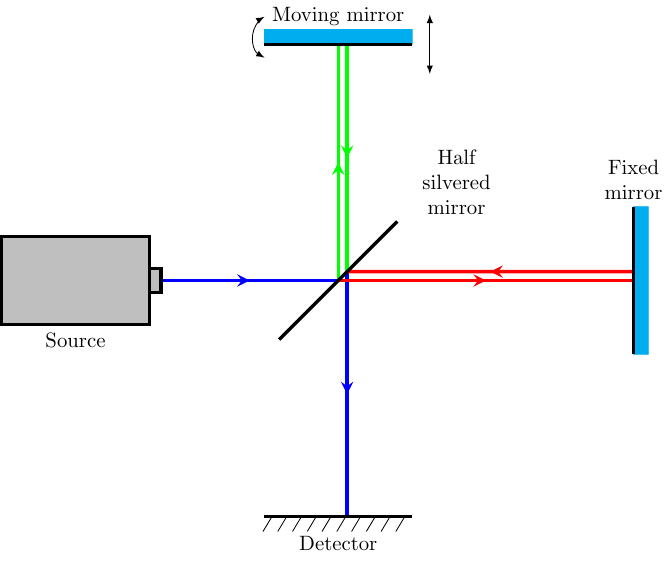}
    \caption{Michelson interferometer}
    \label{fig:michelson}
\end{figure}

The \ac{fts} defines a family of optical devices which measures a spectrum by exploiting the principle of interferometry, that is, by combining a set of coherent waves which travel different paths, whose difference, in term of optical length, is known as the \ac{opd} $\delta$. The raw acquisition of the device, known as \textit{interferogram}, is obtained by varying the \ac{opd} and requires a step of digital processing to retrieve the spectrum in its original domain.

In order to capture an interferogram, the \ac{fts} implements a mechanism to split the incoming light wave into multiple interfering components associated to different traveled paths. Different approaches to achieve this goal allow to distinguish among different classes of \acp{fts}~\cite{Grif07}. 
The Michelson interferometer shown in Figure~\ref{fig:michelson} is the most common example of \ac{fts}. The device uses a half silvered mirror to split the incoming wave into a directly transmitted one and one that is reflected over a second mirror surface. 
The optical path difference between the split waves can be adjusted by modifying the relative distance or angle of one of he reflective mirrors. Assuming no loss of coherency, only a phase shift is introduced among the split waves, and the interference is measured when they recombine over a focal plane.

In mathematical terms, the $k$-th acquisitions $y_k$ associated to the $k$-th \acp{opd} $\delta_k$ can be modeled as \cite{Sale91}:

\begin{equation}
	y_k = \int\limits_0^{B} \mathcal{T}(\sigma, \delta_k)\mathcal{S}(\sigma) \,d\sigma\,,\;\;\;\;k\in[0,..., M-1],
	\label{eq:interferogram}
\end{equation} 
where:
\begin{itemize}
	\item $\mathcal{S}(\sigma)$ denotes the spectrum to reconstruct. For simplicity, we express the spectrum in terms of wavenumbers $\sigma$, the reciprocal of the wavelength;
	\item $B$ is the bandwidth of the instrument in the wavenumber domain;
	\item $\mathcal{T}(\sigma, \delta_k)$ is the transmittance response of the device for a given \ac{opd} $\delta_k$. This transmittance includes various device-specific mechanisms such as reflections, refractions, attenuations, integration time, and so on.
\end{itemize}

The vector $\mathbf{y}=\{y_k\}_{k=[0,...,M-1]}$ can be interpreted as a sampled version of an ideally continuous interferogram, and the goal of the reconstruction is to obtain an estimation $\widehat{\mathbf{x}}$ of the digital representation $\mathbf{x}=\{x_n\}_{n\in[0,..,N-1]}$ of $S(\sigma)$. This representation can be obtained as a numerical integration of the spectrum over $N$ intervals of length $\Delta\sigma=B/N$, that is:

\begin{equation}
	x_n := \frac{1}{\Delta\sigma}\int\limits_{n\Delta\sigma}^{(n+1)\Delta\sigma} \mathcal{S}(\sigma)\;d\sigma\,,
	\label{eq:x}
\end{equation}
where, assuming that $\mathcal{T}$ can be approximated by its value in the midpoint of each interval, yields:
\begin{equation}
	y_k \approx \sum\limits_{n=0}^{N-1} \mathcal{T}\left(\frac{2n+1}{N}\Delta\sigma, \delta_k\right) x_n\,.
	\label{eq:numerical}
\end{equation}

The model described above is relevant for single-pixel acquisitions; various designs have been proposed for imaging systems based on this principle, where the sensor on the focal plane is capable to capture the interferogram associated to different points on the targeted scene. A common example for such cases involves devices based on Fabry-Perot interferometry, as samples of the interferogram is obtained by allowing the light rays to follow different paths within an etalon with parallel faces.
Such designs may for example involve single-interferometers with diffraction gratings~\cite{Kuhn19}, or more commonly multi-aperture snapshot spectrometers where interferometers with different thicknesses are disposed over a staircase matrix~\cite{Oikn18}, and the raw acquisition is composed of several subimages. This image formation model is described in detail in~\cite{Pico23a}, with several algorithmic solutions that have been proposed for optimizing the raw acquisitions~\cite{Joun23}, a subset of which is addressed in this work.

In the following sections, we will focus only on the situation where the acquisition $\mathbf{y}$ represents a single-pixel interferogram; for multi-aperture devices, this is obtained by selecting matching pixels across different subimages. The estimation of the spectrum $\mathbf{x}$ is related to a single point on the scene and obtained by applying the chosen inversion algorithm over a vector $\mathbf{y}$.

\subsection{FFT-based inversion}
\label{ssec:problem_fft}

The family of strategies based on the transformation in the Fourier domain assumes that the transfer function $\mathcal{T}(\sigma, \delta_k)$ of eq.~\eqref{eq:interferogram} is given by a specific analytical expression.

To derive such expression, let us assume that incident spectrum $\mathcal{S}(\sigma)$ is split into two interfering components $\mathcal{S}_0$ and $\mathcal{S}_1$, whose optical intensities are respectively $\mathcal{S}_0(\sigma)=a \mathcal{S}(\sigma)$ and $\mathcal{S}_1(\sigma)=a r \mathcal{S}(\sigma)e^{j\phi}$. Here, the coefficients $a$ and $r$ denote the subsequent attenuation factors of each of the two waves, while $\phi=2\pi\sigma\delta_k$ is a phase shift introduced by the \ac{fts}, which depends on its operating principle.

The interferogram $y_k$ is then given by integrating the collected energy over the whole wavenumber range:

\begin{equation}
	y_k=\int\limits_0^{B}\lvert \mathcal{S}_0+\mathcal{S}_1 \rvert \, d\sigma=a\int\limits_0^{B}\left(1+r \cos(\pi\sigma\delta)\right)\mathcal{S}(\sigma)\,d\sigma\,,
\end{equation}
which can be rewritten as:
\begin{equation}
	\widetilde{y}_{k}:=\frac{\frac{y_k}{a}-\overline{\mathcal{S}}}{2r}=\int\limits_{0}^{B}\cos(\pi\sigma\delta_k)\mathcal{S}(\sigma)\,d\sigma\,,
	\label{eq:fts}
\end{equation}
where $\overline{\mathcal{S}} = \int\limits_0^{B} \mathcal{S}(\sigma)\,d\sigma$ is the average value of input spectrum and $\mathcal{S}(\lvert \sigma \rvert)$ is its even symmetrical extension of $\mathcal{S}(\sigma)$ to negative wavenumbers.
If we assume that $\overline{\mathcal{S}}$ is either known (i.e., measuring it for an \ac{opd} equal to 0) or zero (i.e., if the spectrum is low pass filtered before the measurement), and $a$ and $r$ are known, the inverse Fourier transform of $\widetilde{y}_k$ allows to recover the desired spectrum.
By performing the numerical integration~\eqref{eq:numerical}, eq.~\eqref{eq:fts} becomes:
\begin{subequations}
	\begin{align}
		\widetilde{y}_k &= \sum\limits_{n=0}^{N-1}\cos\left(\frac{\pi (2n+1)}{2N}\Delta\sigma \delta_k\right) x_n\label{eq:dct_irregular}\\
		&=\sum\limits_{n=0}^{N-1}\cos\left(\frac{\pi k(2n+1)}{2N}\Delta\sigma \Delta\delta\right) x_n\,,
		\label{eq:dct_regular}
	\end{align}
	\label{eq:dct}
\end{subequations}

where the second step assumes that the \acp{opd} is regularly sampled with a step size $\Delta\delta$, or in other words $\delta_k = k \Delta\delta$. 
Eq.~\eqref{eq:dct_regular} is equal to the \ac{dct}-II expression up to a scaling factor, assuming $\Delta\sigma \Delta\delta=1$ and $N=M$. Hence, the spectrum reconstruction can be performed with an inverse \ac{dct}. The \ac{dct} can be implemented efficiently with slight variations on the \ac{fft} transformation (i.e. through Makhoul formula~\cite{makhoul1980fast}). However, the interferogram is rarely sampled regularly in practice, which typically demands the interferogram to be interpolated over equally spaced \acp{opd} as pre-processing step.

\subsection{SVD-based methods}
\label{ssec:problem_svd}

In this family of techniques, the model of eq.~\eqref{eq:numerical} is defined as a linear transformation:
\begin{equation}
	\mathbf{y} = \mathbf{A}\mathbf{x}\,.
	\label{eq:direct}
\end{equation}
In the above equation, the transfer matrix $\mathbf{A}\in\mathbb{R}^{M \times N}$ is generally not square and whose coefficients $a_{kl}$ are given by:
\begin{equation}
	a_{kn}:=\mathcal{T}\left(\frac{2n+1}{N}\Delta\sigma, \delta_k\right)\,,\;\;\;\;\;\; \begin{matrix}\forall k\in[0,...,M-1]\\\forall n\in[0,...,N-1]\end{matrix}
\end{equation}

Compared to the previous case, this formulation allows for more generality, as it allows to express any type of linearizable transfer function, e.g. in the case of Fabry-Perot interferometers, this is given by the Airy distribution~\cite{Hari10}:
\begin{equation}
	\mathcal{T}(\sigma, \delta_k) =  \frac{a}{(1-r)^2+4r\sin^2(\pi\delta_k\sigma)}\,,
\end{equation} 
where $a$ is an attenuation factor and $r$ is the reflectivity of the Fabry-Perot etalon, while $\delta_k$ is the \ac{opd} introduced by a single round trip reflection within the cavity.
In the approach of linear regression, the estimation $\widehat{\mathbf{x}}$ of $\mathbf{x}$ is set up as the solution of the following problem:
\begin{equation}
	\widehat{\mathbf{x}} = \arg\min_{\mathbf{x}\in\mathbb{R}^N} \left\|\mathbf{A}\mathbf{x}-\mathbf{y}\right\|_2=\mathbf{A}^{\dagger}\mathbf{y}\,,
	\label{eq:hadamard}
\end{equation}
where $\|\cdot\|_2$ denotes the $\ell_2$ norm and $\mathbf{A}^{\dagger}=(\mathbf{A}^T\mathbf{A})^{-1}\mathbf{A}^T$ is Moore-Penrose pseudo inverse of $\mathbf{A}$.
The pseudo-inversion is efficiently implemented through the \ac{svd} on the matrix $\mathbf{A}$ and then taking the reciprocal of the obtained \acp{sv}~\cite{Stew93}. In other terms, this is a two steps operation:
\begin{align}
	\mathbf{A}&=\mathbf{U}\mathbf{\Xi}\mathbf{V}^T\,,\\
	\mathbf{A}^{\dagger}&=\mathbf{V}\mathbf{\Xi}^{-1}\mathbf{U}^T\,
	\label{eq:svd}
\end{align}
where $\mathbf{U}$ and $\mathbf{V}$ are semi-orthogonal matrices and $\mathbf{\Xi}$ is a diagonal matrix, whose elements on the main diagonal $\{\xi_r\}_{\range{r}{1}{R}}$, are ordered in increasing order. The matrix $\mathbf{\Xi}^{-1}$ is still diagonal and the elements on the main diagonal $\{1/\xi_r\}_{r\in[1,...,R]}$ are the reciprocal of those of $\mathbf{\Xi}$.

Despite its simple mathematical formulation, the closed form solution of eq.~\eqref{eq:hadamard} holds almost no practical value in real scenarios , as the problem is ill-posed or ill-conditioned in the sense of Hadamard. Specifically, the multiplication by $\mathbf{A}^\dagger$ enhances the noise included in the acquisition, causing instability in the results. To avoid this issue, a widespread approach is to employ a modified version $\mathbf{\Xi}'$ of $\mathbf{\Xi}^{-1}$~\cite{Witt09}, whose singular values are penalized. The estimation is hence obtained as:




\begin{equation}
	\widehat{\mathbf{x}}=\mathbf{V}\mathbf{\Xi}'\mathbf{U}^T\mathbf{y}\,,
	\label{eq:pmd}
\end{equation}

The \acp{sv} $\{\zeta_r\}_{r=[1,...,R]}$ of $\mathbf{\Xi}'$ can be chosen with techniques such as the {\ac{tsvd}}~\cite{Hans90} or the \ac{tik}~\cite{Tikh95}, defined as follows:

\begin{subequations}
	\begin{align}
		\zeta_r&=
		\begin{cases}
			1/\xi_r\,,& \textrm{if } 1\leq r\leq R'\,,\\
			0\,,   & \textrm{if } R'<r\leq R\,,
		\end{cases}&\ac{tsvd},\label{eq:tsvd}\\
		\zeta_r&=\frac{\xi_r}{\xi_r^2+\lambda^2}\,,&\ac{tik}.\label{eq:tik}
	\end{align}
	\label{eq:psvd}
\end{subequations}

In the above equations, the matrix rank $1\leq R'\leq R$ and the ridge regression parameter $\lambda\geq0$ are user-defined, and the \ac{pinv} is obtained as a special case when $R'=R$ and $\lambda = 0$, respectively. 
The parameters $R'$ and $\lambda$ are usually chosen according to the amount of noise present in the system~\cite{Gala92}; small values of $R'$ or large values of $\lambda$ introduce a stronger penalization which allows to compensate to smaller \ac{snr} scenarios.
For optical device operating in different conditions (e.g., lighting, temperature, environmental conditions), these parameters have to be set dynamically. For example, they can be selected from a set of possible choices, or be estimated from the characteristics of the acquisition. Techniques such as the L-curve method~\cite{Hans99} have tried to address this problem, but with strong limitations~\cite{Hank96}.
A hardware implementation which is able to provide a quick inversion regardless of the choice of such parameter is an important asset at the consumer's disposal to improve the quality of the reconstructed signals; we aim to address this issue in this paper.

\section{Hardware implementations}
\label{sec:hardware}

This section covers the hardware implementation of the described algorithms. For hardware implementation, the most limiting factor for accurate computation involve computation noise. This noise refers to errors in computation due to hardware imperfections or algorithmic limitations, for which numerical stability is a desirable characteristic of our implementation to assure that those errors do not lead to significant inaccuracies on the final results.
As a priority, we want to ensure that the implemented algorithms properly converge to the desired result with limited distortions.

Additionally, we focus our attention on enhancing performance through data parallelism, achieved by increasing parallel memories. This adaptation enables tuning the architecture for the desired computation performance.


Data parallelism is a technique that involves processing large datasets simultaneously using multiple processing units. Parallel memories, on the other hand, refer to a memory architecture that enables multiple memory modules to work in parallel. To achieve efficient data processing, parallel memories play a vital role as they enable concurrent data access. In this way, the memory system can handle multiple data access requests simultaneously, reducing potential bottlenecks and significantly improving the overall computation performance.

\subsection{Spectrum inversion using FFT}
\label{sec:fft}

The \ac{fft} reconstruction addresses the issue of recovering an estimation of the input $\mathbf{x}$ from regularly sampled interferograms as obtained in eq.~\eqref{eq:dct_regular}.
Despite only working under restricting conditions, it is vastly more efficient to use a \ac{fft} than a generalized \ac{dft} in terms of hardware implementation. 

The \ac{fft} requires to respect a certain order of calculations, since the data at a certain stage depends on previous computations. The Cooley-Tukey radix-2~\cite{cooley1965algorithm} has proven to be a very popular choice to implement the \ac{fft} algorithm, as its memory-based architecture allows to freely choose the order of execution. In this work, we use the \ac{bfp} as data format for improved precision in data representation.

We propose to enhance the existing Cooley-Tukey implementation by introducing a pre- and post-processing normalization step for each computational block. The main idea of this proposal is to share the exponent of the \ac{bfp} precision between \ac{fft} stages, eliminating the drawbacks of the normalization block.
This modification, implemented within the constraints of \ac{hls} which impose to follow specific design templates, results in accelerated calculations about 3 times faster than the classic implementation.

We firstly introduce the \ac{bfp} precision in Section~\ref{ssec:bfp} and we showcase the proposed modifications to the Cooley-Tukey implementation in Section~\ref{ssec:fft_inversion}, detailing its pre- and post-processing normalization steps.

\subsubsection{Block floating point (BFP)}
\label{ssec:bfp} 

In this paper we employ \ac{bfp}  as a data format. This choice allows for improved precision and efficiency in representing the data, as it combines the benefits of \ac{fxp} and \ac{fp} data formats. The \ac{bfp} is a variant of \ac{fxp} format that includes an exponent component within a block of $N$ values ($x_0, \: x_1, \: ... \: x_{N-1}$). The shared exponent determines a scale for the normalized block ($x'_0, \: x'_1, \: ... \: x'_{N-1}$), similar to how FP handles each individual variable \cite{kalliojarvi1996roundoff}:

\begin{equation}\label{eq:bfp_1}
	[x_1, ..., x_N] = [x'_1, ..., x'_N]\cdot 2^\gamma,\;\;\;\;  x'_i=2^{-\gamma}x_i\;.
\end{equation}

In the case of \ac{fft}, a block consists of all the values calculated in a single stage. Therefore, the inputs of an \ac{fft} stage are scaled based on the maximum value from the previous stage. The position of the leading bit in the maximum value determines the magnitude of the scale accordingly.

Subsequently, the next input block is shifted according to the leading bit position helping to prevent overflow while the overall shift value is recorded in the exponent component $\gamma$. Ultimately, the correct scale is derived from the shared exponent in eq. \eqref{eq:bfp_1}. Consequently, \ac{bfp} avoids unnecessary truncation, resulting in an improved dynamic range, while maintaining lower computational overhead \cite{koutsoyannis2012improving, mitra2008finite}.

\subsubsection{Proposed inversion architecture} 
\label{ssec:fft_inversion}

\begin{figure}[!ht]
    \centering
    \includegraphics[width=1.0\columnwidth]{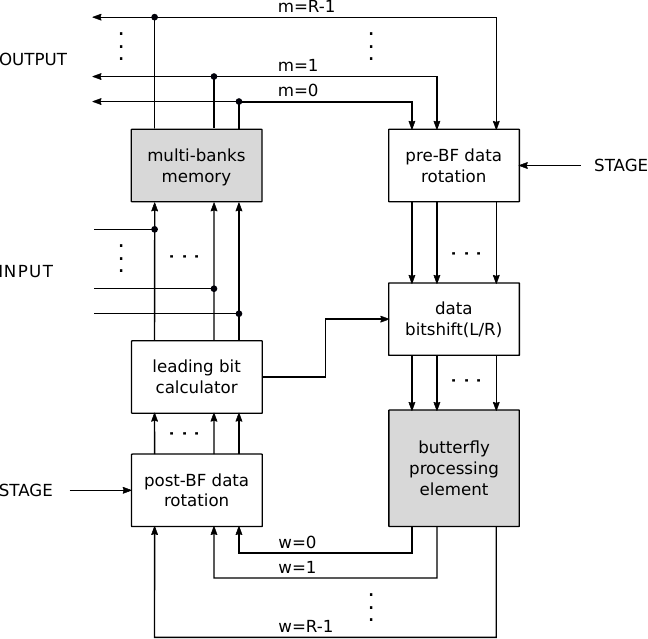}
    \caption{FFT computation structure using parallel memories. The presence of R memory banks enables the calculation of a radix-R \ac{fft} complex butterfly operation in a single clock cycle. For radix-2 scheme, these memory banks provide parallel access to the two complex operands required for each operation in the \ac{fft} stages.
    } 
    \label{fig:fft_main}
\end{figure}

The implemented memory-based architecture is illustrated in Figure~\ref{fig:fft_main}. It features a single butterfly and sequential operations, with required data loaded from memory as the computation progresses. Multi-bank memory is used to enable the calculation of all branches of the butterfly in parallel and the data are initialized in a way that avoids conflicts and then enhance parallelism. In accordance with the \ac{fft} graph \cite{johnson1992conflict}, pre- and post-butterflies swapping schemes are necessary to maintain data separation. Failure to do so after the first \ac{fft} stage can result in required operands being placed in the same memory bank, leading to improper operation. The proposed design incorporates a single butterfly and memory structure with multiple banks, enabling the complex data operands to be loaded in parallel as computation is progressing.

The consistency and regularity across all \ac{fft} stages facilitate to use optimization techniques such as pipelining and unrolling to achieve faster computation. Pipelining with instruction-level parallelism ensures continuous execution across blocks, starting new iterations before previous ones finish~\cite{olukotun1996case} while the initiation interval defines the timing interval between consecutive iterations. Both parameters control the scheduling during \ac{hls} generation. In the case of our design, the \ac{fft} loops are pipelined with an initiation interval of one clock cycle as there are enough ports in the memories to read the data in parallel. Furthermore, the remaining loops responsible for bit operations, including leading bit and address generator, are unrolled to maximize performance and enhance overall efficiency.

\paragraph{Pre-butterfly normalization} Pre-butterfly normalization shifts input data in stage $T$ based on the \textit{leading bit} computed at the end of stage $T-1$. This dependency prevents a scheduling generation with a small initiation interval at high frequencies. With an initiation interval of 1, lower frequencies enable the generation of a scheduling that aligns the leading bit and normalization block within a single cycle, ensuring correct operation. However, at higher operating frequencies, delays in butterfly arithmetic operations exceed a single clock cycle. This delay results in having the leading bit at least one clock cycle ahead of the normalization block in stage $T$ which creates a data feedback loop, hindering the correct schedule generation by \ac{hls} tool. One solution is to increase the initiation interval beyond the feedback delay, which can resolve scheduling problems. However, this extends the computation cycle count.

\paragraph{Post-butterfly normalization} The proposed post-butterfly normalization directly shifts the outputs of the butterfly to the next stage. This modification has an impact on accuracy, as the effect of bit growth in a given stage is corrected at the end of the following stage. To ensure proper operation, it is necessary to select the number of bits for the integer part that can accommodate the worst-case bit growth between two successive stages. Theoretically, for radix-2, the worst-case bit growth is 4 bits. However, in practice, considering the arithmetic growth factor of only 2.41 for one stage as detailed in \cite{elam2003block}, the expected worst-case bit growth between two successive stages is reduced to only 3 bits.

Post-butterfly normalization allows for maximum performance and is hence we propose it as the preferred choice for implementation in this paper.



\subsection{Spectrum inversion using matrix inversion}

In this section, we present hardware strategies for efficient matrix multiplication-based reconstruction. We also explore parallelism enhancements in hardware, focusing on their potential to significantly accelerate computations. 

\subsubsection{Pseudo-inversion (PINV) approach}

The \ac{pinv} strategy involves an inversion without regularization and whose closed form expression is given in eq.~\eqref{eq:hadamard}.

\begin{figure}[!ht]
  \centering
  \includegraphics[width=0.8\columnwidth]{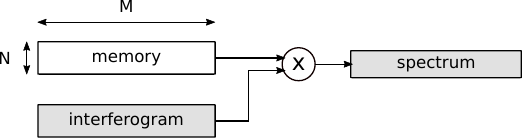}
  \caption{\ac{pinv} computation structure with a single memory.}
  \label{fig:pinv_single_mem}
\end{figure}

The implementation involves matrix multiplication, with the pseudo-inverse matrix $\mathbf{A}^\dagger\in\mathbb{R}^{N \times M}$ and the interferogram vector serving as inputs requiring to conduct $N \times M$ multiplications. To illustrate this process, a simplified structure of a single multiplier is depicted in Figure \ref{fig:pinv_single_mem}. In hardware implementation, additional operations such as memory read and memory write are involved and must be considered in the scheduling process, resulting in a latency for the generated architecture that exceeds the $N \times M$ theoretical latency. Pipelining, unrolling and data parallelism can help to significantly alleviate the latency.

\begin{figure}[!ht]
  \centering
  \includegraphics[width=1\columnwidth]{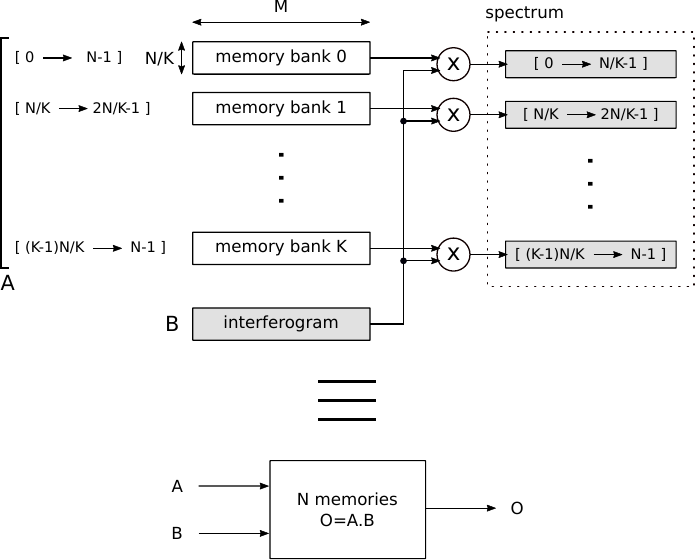}
  \caption{PINV parallel computation structure with multiple memories.}
  \label{fig:pinv_mult_mem}
\end{figure}

Multiple parallel line-column multiplications can drastically speed up computation since each output element's multiplication is independent. Yet, using just one memory with the inversion matrix (Figure \ref{fig:pinv_single_mem}) is not feasible due to the two memory port limitation. To address this, we can split the rows of the matrix $\mathbf{A}^\dagger$ into $K$ subsets and store them in separate memories (Figure \ref{fig:pinv_mult_mem}). This allows to accelerate the computation by a factor of $K$ compared to the single-memory setup. This new structure presents $K$ inversions, each with a transfer matrix having $(N/K)$ rows and $M$ columns. Hardware optimizations from the single memory version apply to theses new inversions as well. The parameter $K$ controls the computational efficiency of the produced implementations and enables assessment of both parallelism's advantages and its effect on hardware resources.

\subsubsection{Penalized SVD approach}

As discussed in Section~\ref{ssec:problem_svd}, both the \ac{tsvd} and \ac{tik} methods are inversion techniques consisting in penalizing the singular values contained in the diagonal matrix $\mathbf{\Xi}'$ obtained by \ac{svd} of the pseudo-inverse $\mathbf{A}^\dagger$, as shown in eq.~\eqref{eq:pmd}. We discuss in this section their hardware implementation.

\begin{figure}[!ht]
  \centering
  \includegraphics[width=1\columnwidth]{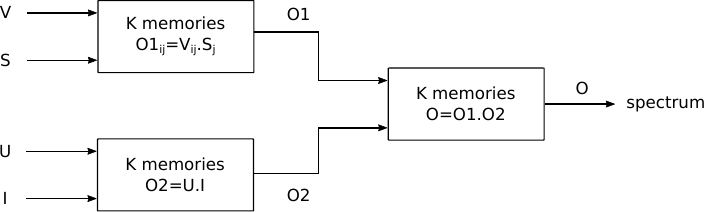}
  \caption{TSVD/TIK parallel computation structure with multiple physical memories (separated ports for maximum data throughput) }
  \label{fig:svd_mult_mem}
\end{figure}

In practical terms, eq.~\eqref{eq:pmd} is composed by three matrix-vector multiplications that share the same computation scheme: $\mathbf{O}_1 = \mathbf{V} \mathbf{\Xi}'$, $\mathbf{O}_2 = \mathbf{U}^T \mathbf{y}$, and $\widehat{\mathbf{x}} = \mathbf{O}_1 \mathbf{O}_2$. 
The term $\mathbf{O}_1$ is independent on $\mathbf{y}$, so it is not required to be computed every time. However, the regularization parameter (i.e., $R'$ for the \ac{tsvd} and $\lambda$ for the \ac{tik}) may change according to the specific conditions of the acquisitions and across different pixels. We hence implement the worst case scenario in order to be able to keep the adaptability of the parameters for every acquisition. Therefore, the implemented calculation structure is using multiple, physically distinct memories so that the parallelism can be applied through multiple memories for each matrix multiplication, as illustrated in Figure~\ref{fig:svd_mult_mem}.


For a matrix of size $N \times M$, the total number of multiplication-addition operations in \ac{svd}-based inversions is given by $R'(2N+M)$, where $R'$ denotes the number of singular values considered in truncation. In case of \ac{tik}, $R'$ is equal the total number of singular values. Further improvements in latency can be achieved by introducing more parallelism, similar to the \ac{pinv} method. 
The parallelism parameter $K$ remains the same for all the multiplication blocks, and the structure is similar for both the \ac{tsvd} and \ac{tik} inversion methods.

\section{Results and discussion}
\label{sec:experiments}


The purpose of this section is to conduct a comprehensive evaluation of hardware implementation strategies for reconstruction. The focus will be on assessing inversion speed, parallelism, and reconstruction quality, in order to gain valuable insights into the performance and effectiveness of these strategies.
We provide a brief summary of the experimental setup, highlighting the key components and configurations used. Subsequently, the discussion focuses on the resource overhead associated with the implementation and the inversion speed of the considered inversion architectures. After that, the role of parallelism is investigated to assess its impact on computation speed and resource utilization for each of the considered methods. Finally, the achieved reconstruction quality is evaluated, providing insights into the accuracy of reconstruction based on the method and architecture parameters. These analyses aim at providing a comprehensive understanding of hardware implementation strategies for reconstruction and present informed observations and conclusions regarding their performance.

\subsection{Experimental setup}


All the considered inversion algorithms have been implemented for Xilinx Zynq SoC FPGA. The experiments have been carried out on a Zybo Z7 dev board, which is equipped with a Xilinx Zynq-7020 SoC (XC7Z020-1CLG400C). The board includes an FPGA Fabric with 85,000 logic elements, 220 DSP slices and 4.9 Mb of memory blocks. Furthermore the chip includes some software/Hardware co-design elements such as a 667 MHz ARM Cortex-A9 processor and 1GB DDR3 RAM. The considered implementations lied purely on the FPGA fabric.
The RTL has been generated with Catapult \ac{hls} IDE while the synthesis have been performed on Vivado for the selected part.

\subsection{Discussion on resource overheads}

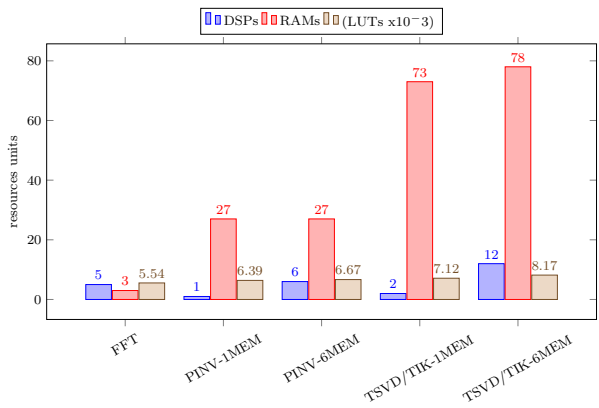
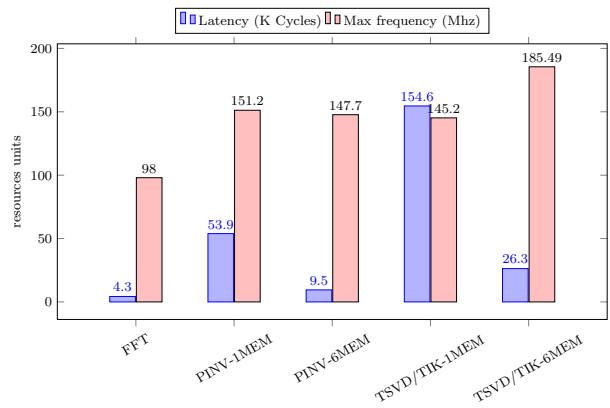
\begin{figure*}[!ht]
    \centering
    \begin{subfigure}[b]{0.49\textwidth}
        \centering
\begin{tikzpicture}[scale=0.6]
    \begin{axis}[
    ybar = 0.6,
    width=1.6\columnwidth,
    height=0.9\columnwidth,
    enlarge x limits = 0.2,
    bar width=16pt,
    legend style={at={(0.5,1.13)},
    anchor=north,legend columns=0},
    ylabel={resources units},
    symbolic x coords={FFT, PINV-1MEM, PINV-6MEM, TSVD/TIK-1MEM, TSVD/TIK-6MEM},
    xtick=data,
    xticklabel style={rotate=30},
    nodes near coords,
    ]
    \addplot coordinates {(FFT,5) (PINV-1MEM,1) (PINV-6MEM,6) (TSVD/TIK-1MEM,2) (TSVD/TIK-6MEM,12)}; 
    \addplot coordinates {(FFT,3) (PINV-1MEM,27) (PINV-6MEM,27) (TSVD/TIK-1MEM,73) (TSVD/TIK-6MEM,78)}; 
    \addplot coordinates {(FFT,5.54) (PINV-1MEM,6.39) (PINV-6MEM,6.670) (TSVD/TIK-1MEM,7.12) (TSVD/TIK-6MEM,8.17)}; 

    \legend{DSPs, RAMs, (LUTs x$10^-3$)}
    \end{axis}

\end{tikzpicture}
\caption{Hardware resources comparison.} 
\label{fig:resources_overhead}
    \end{subfigure}
    \hfil
    \begin{subfigure}[b]{0.49\textwidth}
        \centering
\begin{tikzpicture}[scale=0.6]

    \begin{axis}[
    ybar = 0.6,
    width=1.6\columnwidth,
    height=0.9\columnwidth,
    enlarge x limits = 0.2,
    bar width=16pt,
    legend style={at={(0.5,1.13)},
    anchor=north,legend columns=0},
    ylabel={resources units},
    symbolic x coords={FFT, PINV-1MEM, PINV-6MEM, TSVD/TIK-1MEM, TSVD/TIK-6MEM},
    xtick=data,
    xticklabel style={rotate=30},
    nodes near coords,
    ]
    \addplot coordinates {(FFT,4.3) (PINV-1MEM,53.9) (PINV-6MEM,9.5) (TSVD/TIK-1MEM,154.6) (TSVD/TIK-6MEM,26.3)}; 
    \addplot[fill=pink] coordinates {(FFT,98) (PINV-1MEM,151.2) (PINV-6MEM,147.7) (TSVD/TIK-1MEM,145.2) (TSVD/TIK-6MEM,185.49)}; 

    \legend{Latency (K Cycles), Max frequency (Mhz)}
    \end{axis}

\end{tikzpicture}
\caption{Latency / Max frequency comparison.}
\label{fig:lat_freq_comp}
    \end{subfigure}
    \caption{Hardware resources and latency comparison results. 
    }
    \label{fig:freq_comp}
\end{figure*}

The comparison of hardware resources is presented in Figure~\ref{fig:resources_overhead}. Considering DSP blocks without parallelism, the matrix multiplication-based methods (\ac{pinv}, \ac{tsvd}, and \ac{tik}) exhibit lower overhead. The \ac{pinv} method requires a single multiplier for the matrix-vector multiplication, while the \ac{svd}-based methods (\ac{tsvd} and \ac{tik}) utilize two multipliers for the $\mathbf{V} \mathbf{\Xi}'$ and $\mathbf{U}^T\mathbf{y}$ multiplications. These multiplications are computed sequentially, using one multiplier for each operation. On the other hand, the \ac{fft} method requires 5 multipliers since all the operations are performed in a complex form.

In terms of memory usage, the \ac{fft} method proves to be more advantageous with the lowest overhead. Only the twiddle factors and temporary data need to be stored in memory for this method. In contrast, the matrix multiplication-based methods require a larger number of memory blocks. Specifically, the \ac{pinv} method requires memory to store the inversion matrix, while the \ac{tsvd}/\ac{tik} methods require memory for storing the elements of the \ac{svd} factorization.

\subsection{Discussion on inversion speed}



Computation speed is achieved by dividing the operation count by the maximum operating frequency. Parallelism is dictated by available memory ports, and it affects the speed through the parallelism factor $K$. Figure~\ref{fig:lat_freq_comp} illustrates operating performances. Among methods, the \ac{fft} exhibits the swiftest performance, measured as clock cycles divided by max frequency. This surpasses non-parallel matrix multiplication approaches.

It is approximately 8 times faster than \ac{pinv} and 24 times faster than \ac{tik} with a single memory (Figure~\ref{fig:lat_freq_comp}). These improved performances can be attributed to the reduced redundancy in the \ac{fft} algorithm. However, the speed difference can be mitigated by utilizing data parallelism. For example, by employing 6 memories in parallel for \ac{pinv} and \ac{tik}, the processing speed is only 1.5 and 3.2 times slower, respectively, compared to \ac{fft}. Further reduction in the speed difference can be achieved by increasing the number of memories, albeit at the expense of additional DSP blocks. In the case of \ac{pinv}, the number of DSP blocks increases proportionally to the number of memories, while this rate doubles for \ac{tik}.

The reported speed for \ac{tsvd} corresponds to the worst case scenario where all the singular values are considered. However, when a lower number of singular values is used, \ac{tsvd} can achieve improved inversion speed compared to \ac{pinv}. On the other hand, \ac{tik} consistently performs 4 times slower than \ac{pinv} since all the singular values are utilized in the calculation.


\subsection{Discussion on parallelism in \ac{tsvd}/\ac{tik}}

As discussed earlier, both \ac{tsvd} and \ac{tik} methods share the same computing architecture. Unlike the \ac{pinv} method, this architecture requires additional storage capacity for the three matrices. The overhead resulting from this storage requirement is evident in Figure~\ref{fig:resources_overhead}, where the number of memory blocks used is nearly three times higher compared to \ac{pinv}. This overhead remains relatively constant regardless of the parallelism factor. It is worth noting that the relationship between computational latency and the parallelism factor follows a similar trend as in the case of \ac{pinv}, with a latency factor of 5.5 when the parallelism factor ranges from 1 to 6. From a hardware overhead perspective, \ac{pinv} clearly offers advantages. However, the subsequent discussion will demonstrate that this may not be the case when considering the reconstruction quality.

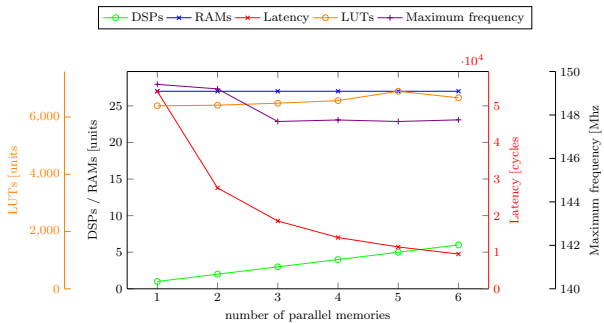
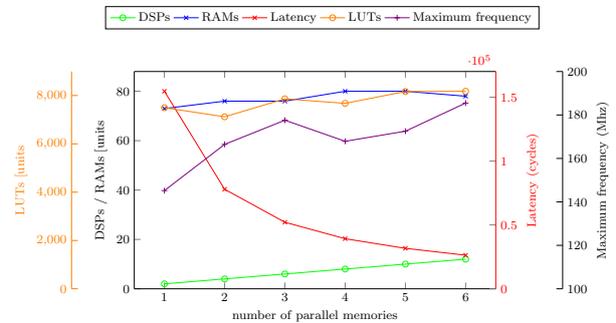
\begin{figure*}[!ht]
    \begin{subfigure}[b]{0.49\textwidth}
        \centering
\begin{tikzpicture}[scale=0.55]
    \pgfplotsset{every axis/.style={ymin=0}}
    \begin{axis}[width=1.2\columnwidth, height=0.8\columnwidth, axis y line*=left, xlabel=number of parallel memories, ylabel=DSPs / RAMs [units]]
        \addplot[green, mark=o, draw] coordinates {(1,1.0)  (2,2.0) (3,3.0) (4,4.0)  (5,5.0)  (6,6.0)};\label{label_1}
        \addplot[blue, mark=x, draw] coordinates {(1,27.0)  (2,27.0) (3,27.0) (4,27.0)  (5,27)  (6,27)};\label{label_2}
    \end{axis}

    \begin{axis}[width=1.2\columnwidth, height=0.8\columnwidth, red, axis y line*=right, axis x line=none, ylabel=Latency [cycles]]%
      
        \addplot[red, mark=x] coordinates {(1,53965)  (2,27559) (3,18529) (4,14014)  (5,11434)  (6,9499)};\label{label_3}
    %
    \end{axis}
    
    \begin{axis}[width=1.2\columnwidth, height=0.8\columnwidth, orange, axis y line*=left,axis x line=none, ylabel=LUTs [units]]
    
      \pgfplotsset{every outer y axis line/.style={xshift=-1.5cm}, every tick/.style={xshift=-1.5cm}, every y tick label/.style={xshift=-1.5cm} }
        \addplot[orange ,mark=o] coordinates {(1,6390)  (2,6413) (3,6483) (4,6572)  (5,6899)  (6,6670)};\label{label_4}
    
    \end{axis}

    \begin{axis}[width=1.2\columnwidth, height=0.8\columnwidth, ymin=140,ymax=150, axis y line*=right, axis x line=none, legend style={at={(0.52,1.3)},
    anchor=north,legend columns=0},ylabel=Maximum frequency [Mhz]]%
    \addlegendimage{/pgfplots/refstyle=label_1}\addlegendentry{DSPs}
    \addlegendimage{/pgfplots/refstyle=label_2}\addlegendentry{RAMs}
    \addlegendimage{/pgfplots/refstyle=label_3}\addlegendentry{Latency}     
    \addlegendimage{/pgfplots/refstyle=label_4}\addlegendentry{LUTs}   
    \pgfplotsset{every outer y axis line/.style={xshift=1.6cm}, every tick/.style={xshift=1.6cm}, every y tick label/.style={xshift=1.6cm} }
        \addplot[violet ,mark=+] coordinates {(1,149.41)  (2,149.2) (3,147.7) (4,147.77)  (5,147.7)  (6,147.776)};\addlegendentry{Maximum frequency}
    \end{axis} 

\end{tikzpicture}
\caption{PINV}
\label{fig:pvin_parallel_mem_comp}
    \end{subfigure}
    \hfil
    \begin{subfigure}[b]{0.49\textwidth}
        \centering
\begin{tikzpicture}[scale=0.55]
\pgfplotsset{every axis/.style={ymin=0}}
    \begin{axis}[width=1.2\columnwidth, height=0.8\columnwidth, axis y line*=left, xlabel=number of parallel memories, ylabel=DSPs / RAMs [units]]
        \addplot[green, mark=o, draw] coordinates {(1,2.0)  (2,4.0) (3,6.0) (4,8.0)  (5,10.0)  (6,12.0)};\label{label_5}
        \addplot[blue, mark=x, draw] coordinates {(1,73.0)  (2,76.0) (3,76.0) (4,80.0)  (5,80)  (6,78)};\label{label_6}
    \end{axis} 
    \begin{axis}[width=1.2\columnwidth, height=0.8\columnwidth, red, axis y line*=right, axis x line=none, ylabel=Latency (cycles)]%
      
        \addplot[red, mark=x] coordinates {(1,154673)  (2,77918) (3,52118) (4,39218)  (5,31693)  (6,26318)};\label{label_7}
    
    \end{axis}
    
    \begin{axis}[width=1.2\columnwidth, height=0.8\columnwidth, orange, axis y line*=left,axis x line=none, ylabel=LUTs [units]]
    
      \pgfplotsset{every outer y axis line/.style={xshift=-1.5cm}, every tick/.style={xshift=-1.5cm}, every y tick label/.style={xshift=-1.5cm} }
        \addplot[orange ,mark=o] coordinates {(1,7496)  (2,7120) (3,7862) (4,7676)  (5,8165)  (6,8176)};\label{label_8}
    
    \end{axis}

    \begin{axis}[width=1.2\columnwidth, height=0.8\columnwidth, ymin=100,ymax=200, axis y line*=right, axis x line=none, legend style={at={(0.52,1.3)},
    anchor=north,legend columns=0},ylabel=Maximum frequency (Mhz)]%
    \addlegendimage{/pgfplots/refstyle=label_5}\addlegendentry{DSPs}
    \addlegendimage{/pgfplots/refstyle=label_6}\addlegendentry{RAMs}
    \addlegendimage{/pgfplots/refstyle=label_7}\addlegendentry{Latency}  
    \addlegendimage{/pgfplots/refstyle=label_8}\addlegendentry{LUTs}         
    \pgfplotsset{every outer y axis line/.style={xshift=1.6cm}, every tick/.style={xshift=1.6cm}, every y tick label/.style={xshift=1.6cm} }
        \addplot[violet ,mark=+] coordinates {(1,145.2)  (2,166.47) (3,177.58) (4,167.89)  (5,172.53)  (6,185.49)};\addlegendentry{Maximum frequency}
    \end{axis} 

\end{tikzpicture}
\caption{TSVD.} 
\label{fig:tsvd_parallel_mem_comp}
    \end{subfigure}
    \caption{Hardware overhead comparison in terms of parallel memories.}
\end{figure*}

\subsection{Discussion on the reconstruction quality}

\begin{figure*}
    \begin{subfigure}[b]{0.49\textwidth}
        \centering
\begin{tikzpicture}[scale=0.8]
    \begin{axis}[
                 width=1\columnwidth,
                 height=0.9\columnwidth,
                 xlabel=Interferogram precision [bits]],
                 ylabel=SNR [dB],
                 legend style={draw=none, font=\footnotesize},
                 legend style={at={(0.8,0.5)}, anchor=north},
                 legend cell align={left},
                 legend entries={{PINV}, {TSVD}, {FFT}, {TIK}}]

                 \addplot [color=blue, mark=square, line width=1pt, mark options={scale=1.1}] coordinates{ (4,13.15)  (6,46.15) (8,73.81) (10,76.05)  (12,77.41)  (14,77.14)  (16,77.08) (18,77.09)};                
                 \addplot [color=black!40!green, mark=o, line width=1pt, mark options={scale=1.1}] coordinates{ (4,13.91)  (6,57.24) (8,81.33) (10,81.98)  (12,81.81)  (14,81.79)  (16,81.77)  (18,81.76)};               
                 \addplot [color=orange, mark=triangle, line width=1.2pt, mark options={scale=1.3}] coordinates{ (4,2.17)  (6,8.33) (8,8.08) (10,7.91)  (12,7.96)  (14,7.92)  (16,7.92)  (18,7.99)};
                 \addplot [color=red, mark=o, line width=1.2pt, mark options={scale=1.3}] coordinates{ (4,11.63)  (6,11.27) (8,13.57) (10,38.77)  (12,87.84)  (14,111.58)  (16,114.93)  (18,114.94)};


    \end{axis}
\end{tikzpicture}
\caption{Quality of inversion comparison.} 
    \label{fig:snr_comp}
    \end{subfigure}
    \hfil
    \begin{subfigure}[b]{0.49\textwidth}
        \input{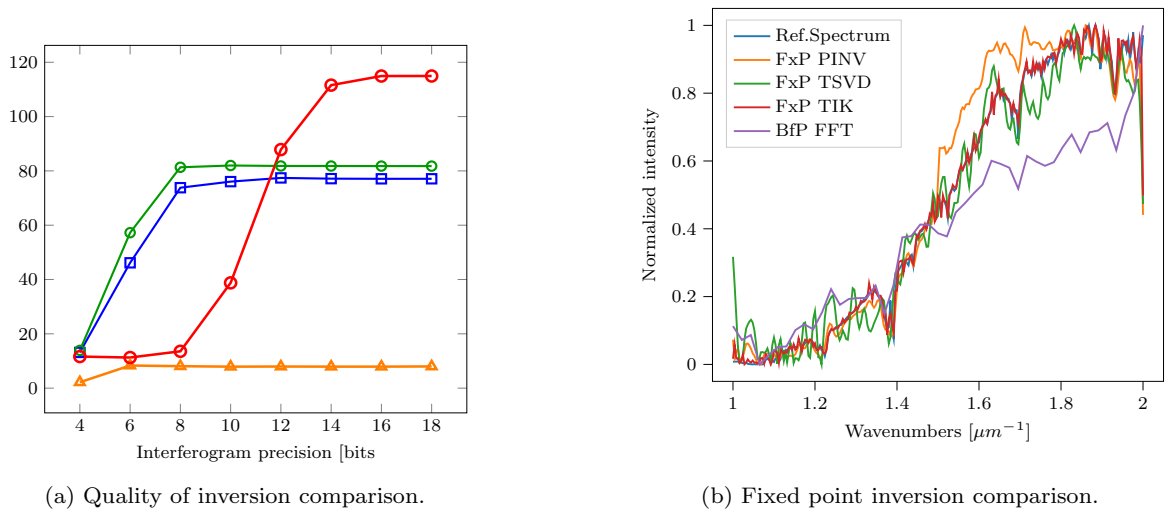}
    \end{subfigure}
    \caption{Comparison of reconstruction quality.}
\end{figure*}

This section focuses on reconstruction quality from different inversion methods, shown in Figure \ref{fig:snr_comp}. \ac{fft} does not guarantee superior quality even with precise interferograms. On the other hand, both \ac{pinv} and \ac{tsvd} methods demonstrate better overall quality, with comparable results regarding the required precision of interferograms to achieve maximum quality. This can be attributed to the fact that \ac{pinv} is a special case of \ac{tsvd}, where all singular values are considered. 

To achieve higher quality reconstructions, the \ac{tik} method proves advantageous as it offers significant improvements over other matrix multiplication-based methods. However, to achieve the same maximum quality as \ac{pinv}/\ac{tsvd}, \ac{tik} requires additional bits (8 bits for \ac{pinv}/\ac{tsvd} vs 12 bits for \ac{tik}). Nevertheless, with 4 bits precision increase, \ac{tik} enables a significant improvement in quality, surpassing the other methods based on matrix multiplication by approximately 30\%. Overall, these findings emphasize the importance of considering reconstruction quality alongside speed and hardware resources when selecting an inversion method.

\section{Conclusion}

In this paper, we conducted a comprehensive evaluation of hardware implementation strategies for spectrum reconstruction from interferograms acquired by FTS.
Specifically, this study focused on inversion speed, parallelism, and reconstruction quality. The resource overhead analysis highlighted the advantages of matrix multiplication-based approaches in terms of \ac{dsp} block utilization and memory requirements. The investigation into inversion speed demonstrated improved efficiency for the \ac{fft} method while parallel matrix multiplication methods exhibited competitive performance at the cost of higher resource utilization.

Considering the reconstruction quality, the \ac{fft} exhibits the lowest reconstruction quality, even with precise interferograms, compared to \ac{pinv} and \ac{tsvd}. Notably, the \ac{tik} method showed promising potential for achieving higher quality reconstructions when considering a slight increase in bit precision. Although \ac{tsvd}/\ac{tik} have architectural similarities, the reconstruction quality assessment revealed nuanced differences between the methods, emphasizing the significance of considering precision requirements and bit-depth when targeting higher-quality results.

By providing a comprehensive understanding of these implementation strategies, this paper intends to facilitate informed decision-making in selecting the most suitable approach for specific applications and target hardware. 

\section*{Conflict of interest}

The authors declare that they have no conflict of interest.

\printbibliography

\end{document}